\documentclass[npg]{copernicus}

\begin{document}

\title{Joule Heating and Anomalous Resistivity in the Solar Corona}

\author[]{Steven R. Spangler}

\affil[]{Department of Physics and Astronomy, University of Iowa, Iowa City, IA 52242}


\runningtitle{Currents in Solar Corona}

\runningauthor{Steven R. Spangler}

\correspondence{Steven R. Spangler\\ steven-spangler@uiowa.edu}

\received{}
\pubdiscuss{} 
\revised{}
\accepted{}
\published{}


\firstpage{1}

\maketitle

\begin{abstract}
Recent radioastronomical observations of Faraday rotation in the solar corona can be interpreted as evidence for coronal currents, with values as large as $2.5 \times 10^9$ Amperes \citep{Spangler07}. These estimates of currents are used to develop a model for Joule heating in the corona. It is assumed that the currents are concentrated in thin current sheets, as suggested by theories of two dimensional magnetohydrodynamic turbulence. The Spitzer result for the resistivity is adopted as a lower limit to the true resistivity.  The calculated volumetric heating rate is compared with an independent theoretical estimate by \cite{Cranmer07}. This latter estimate accounts for the dynamic and thermodynamic properties of the corona at a heliocentric distance of several solar radii.  Our calculated Joule heating rate is less than the Cranmer et al estimate by at least a factor of $3 \times 10^5$. The currents inferred from the observations of \cite{Spangler07} are not relevant to coronal heating unless the true resistivity is enormously increased relative to the Spitzer value. However, the same model for turbulent current sheets used to calculate the heating rate also gives an electron drift speed which can be comparable to the electron thermal speed, and larger than the ion acoustic speed. It is therefore possible that the coronal current sheets are unstable to current-driven instabilities which produce high levels of waves, enhance the resistivity and thus the heating rate.   
\end{abstract}

\introduction

In a recent paper, \cite{Spangler07} reported radioastronomical observations which were consistent with the presence of coronal currents in the range of hundreds of MegaAmperes to a few GigaAmperes.  This measurement was made using Faraday rotation observations of a radio source occulted by the corona, and the coronal plasma probed was at heliocentric distances of 5.2 to 6.7 $R_{\odot}$.  In the present paper, I discuss the implications of these observations for the process of coronal heating by Joule heating.  

As discussed in \cite{Spangler07} the currents reported (and summarized in Section 2 below)  correspond to the {\em net} current within an Amperian loop defined by the two, closely-spaced lines of sight through the corona to the different parts of the radio source.  The measured net current could be, and probably is, a residual due to numerous current filaments with alternate positive and negative current density within the Amperian loop.  

This topic is of interest because Joule heating has been identified as the primary mechanism for heating the closed-field part of the corona \citep{Gudiksen05,Peter06}.  The purpose of this paper is to make model-dependent estimates of the heating rate due to Joule dissipation of these currents.  As expected, the calculation involves introduction of several ``imponderables'', i.e. physical characteristics of the turbulence in the corona which are poorly constrained by observations, but which play an important role in coronal heating.  I feel this exercise is worthwhile in identifying coronal parameters which are important in coronal heating, so that they can be targeted for future observational investigations.   

The outline of this paper is as follows.  In Section 2, I briefly summarize the observational results of \cite{Spangler07} which were the basis of the estimates of the coronal current.  Section 3 is the most important part of the paper; it introduces a model for current-carrying coronal turbulence, and identifies the most important characteristics of this turbulence. A glossary of the variables and parameters introduced in this discussion is given in Table 1. This model is used to obtain an estimate of the volumetric heating rate due to Joule heating. Section 4 briefly considers the possibility that current densities in these sheets could be large enough to generate turbulence via current-driven instabilities, and thus produce high wave levels which enhance the resistivity and thus the Joule heating rate.  Section 5 summarizes what has been learned from this exercise and presents conclusions.  

\section{Brief Summary of Radioastronomical Measurements of Coronal Currents}
The result reported by \cite{Spangler07} was of a difference in the Faraday rotation measure $\Delta RM$ between two lines of sight to two components of an extragalactic radio source (3C228). These lines of sight were separated by an angular distance $\theta$, which corresponds to a linear separation in the corona between the two lines of sight, $l = \theta d$, where $d$ is the distance to the Sun.  In the observations reported in \cite{Spangler07} $\theta = 46$ arcseconds and $l=33,000$ km.  The observations were made when the line of sight to 3C228 passed through the corona at heliocentric distances from $5.2 - 6.7 R_{\odot}$.  The technique is illustrated in Figure 1 of \cite{Spangler07}.  

The fundamental physical relation used in the technique is expressed by equation (3) of \cite{Spangler07}
\begin{equation}
\Delta RM = C \oint n \vec{B} \cdot \vec{ds} \simeq C \bar{n}\oint \vec{B} \cdot \vec{ds}
\end{equation}
where the integral is around an Amperian loop through the corona, consisting of the two lines of sight, closed by imaginary line segments which join the two lines of sight, at locations infinitely separated from the corona.  In this formula, $C$ is a collection of atomic constants which arise in the description of Faraday rotation, defined as $C = \frac{e^3}{8 \pi^2 c^3 \epsilon_0 m_e^2} = 2.631 \times 10^{-13}$ in SI units,  $n(\vec{x})$ is the electron density, $\vec{B}(\vec{x})$ is the vector magnetic field in the corona, and $\vec{ds}$ is an incremental step around the Amperian Loop.  

Use of Ampere's Law 
\begin{equation}
\oint \vec{B} \cdot \vec{ds} = \mu_0 I
\end{equation}
in equation (1) shows that the differential rotation measure $\Delta RM$ is directly related to the current within the Amperian loop defined by the two lines of sight. 
The transition from the middle to the right term in equation (1) involves an approximation, in which the position-dependent plasma density in the integrand is replaced by an effective mean density $\bar{n}$.  This approximation is discussed at length in \cite{Spangler07}, where arguments for its plausibility are presented.  

As is the case with Ampere's Law, the differential Faraday rotation measurement provides information on the {\em net current}  within the Amperian Loop.  In general, we expect both positive and negative currents to be flowing within the loop.  The electrical current flowing in a given location in the corona could be much larger than that deduced by the arguments above, and contained in equation (7) of \cite{Spangler07}.  

\cite{Spangler07} presented the following results from two observing sessions with the Very Large Array\footnote{The Very Large Array is an instrument of the National Radio Astronomy Observatory.  The NRAO is a facility of the National Science Foundation, operated under cooperative agreement with Associated Universities, Inc.} radiotelescope. Each session lasted approximately 8 hours. 
\begin{enumerate}
\item In one of the two sessions, there was a confident detection of a $\Delta RM$ event, with a corresponding inferred electrical current of $2.5 \times 10^9$ Amperes.  
\item In the second observing session, a marginal $\Delta RM$ event was detected with a value of $I$ (which may well be considered an upper limit) of $2.3 \times 10^8$ Amperes. 
\item During a several hour period of good data quality, no significant  $\Delta RM$ events were detected, with a corresponding upper limit to the current of  $7.7 \times 10^8$ Amperes.  
\item Although the data from the earlier investigation of \cite{Sakurai94} have not been reanalysed in this manner, examination of Figure 11 of that paper shows no clear evidence of a $\Delta RM$ event in several more hours of VLA observation.  All of this indicates that detection of a clear differential rotation measure event between lines of sight separated by $\simeq 30,000$ km is relatively rare.  
\end{enumerate} 
The aforementioned observations will represent the observational constraints imposed on the theory developed in the next section.  

\begin{figure}[t]
\vspace*{2mm}
\includegraphics[width=8.3cm]{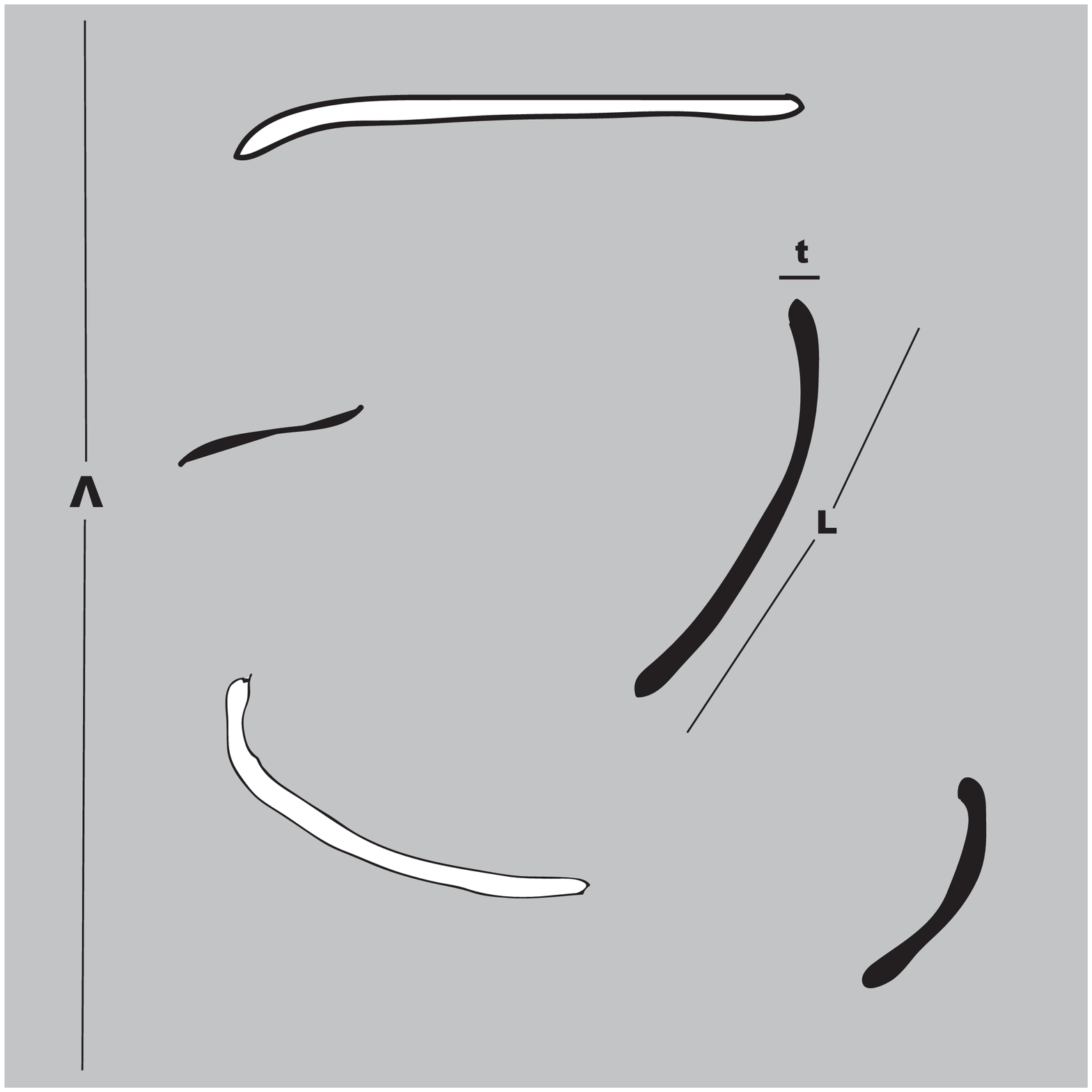}
\caption{A vision of current-carrying coronal turbulence. The 2 dimensional plane represented is perpendicular to the radial direction, which is also the direction of the large scale coronal magnetic field. The figure portrays the current density in gray-scale format. Black represents large positive current density, white is large negative current density, and gray indicates zero current density.  This diagram illustrates the basic model for the current in the corona.  It is contained within intense, narrow current sheets of both positive and negative sign.  There are a few such current sheets, with thickness $t_c$ and width $L_c$ (noted in the figure as ``t'' and ``L''), within a ``domain'' of width $\Lambda$. Adapted from \cite{Spangler99}.}
\end{figure}

\section{Implications for Coronal Heating}
In this section, I discuss the implications of the results from \cite{Spangler07} for coronal heating.  The presence of electrical currents indicates that Joule heating will occur as well.  I will calculate an estimate of the average volumetric heating rate of a system of currents which could produce the observations discussed in Section 2.  

This calculation will be highly model dependent, as well as dependent on assumptions regarding the nature of the current sheets.  Since the subsequent discussion will introduce many assumed 
parameters of the coronal current sheets, some of which are poorly constrained, I will follow the once-common practice in physics and astronomy literature of including a glossary of physical variables.  This is contained in Table 1.  
\begin{table*}[t]
\caption{Glossary of Current Sheet Characteristics}
\vskip4mm
\centering
\begin{tabular}{ll}
\tophline
Variable & Definition  \\
\middlehline
$A$ & Area of the Amperian Loop defined by the experiment\\
$\beta$ & ratio of current sheet width to domain size, $L_c/\Lambda$ \\
$\dot{\cal E}$ & Joule heating per current sheet (Watts)\\
$\dot{E}$ & Joule heating per domain, due to all current sheets \\
$\epsilon$ & Domain-averaged volumetric heating rate\\
$\epsilon_T$ & Domain-averaged volumetric heating rate for turbulent current sheets\\
$\epsilon_S$ & Domain-averaged volumetric heating rate for deterministic current sheets\\
$\eta$ & Electrical resistivity of a plasma, $=1/\sigma$ \\
$\eta_S$ & Spitzer resistivity (equation (16))\\
$f_+$ & Fraction of sheets with positive current density within Amperian Loop \\
$f_-$ & Fraction of sheets with negative current density within Amperian Loop \\
$I$ & Electrical current per sheet\\
$I_{obs}$ & Total current within the Amperian Loop\\
$j$ & Current density within one current sheet \\
$l$ & Transverse separation of two  lines of sight in the corona\\
$L_c$ & Width of a current sheet  \\
$\Lambda$ & Transverse scale of a domain which contains a few current sheets  \\
$\Lambda_z$ & Extent of domain along the large scale magnetic field \\
$\mu $ & Anisotropy of the current sheets (scale length along magnetic field/scale length across field)\\
$N$ & Number of current sheets within one domain \\
$N_+$ & Number of sheets with positive current density within Amperian loop \\
$N_-$ & Number of sheets with negative current density within Amperian loop \\
$N_T$ & Total number of current sheets within the Amperian loop defined by the experiment\\
$S_{los}$ & Effective line-of-sight depth of the coronal plasma \\
$\sigma$ & electrical conductivity of a plasma\\
$t_c$ & Thickness of a current sheet \\
$v_D$ & electron drift speed \\
$V$ & Volume of a current sheet \\
$V_D$ & Volume of a domain \\
$Z_c$ & Extent of a current sheet along the large scale magnetic field \\
\bottomhline
\end{tabular}
\end{table*}
The following analysis assumes that the current is contained in a number of thin current sheets within the Amperian Loop. This model is illustrated in Figure 1. A coordinate system is defined by having one axis (the $z$ axis) coincide with that of the large scale coronal magnetic field. I assume that the current sheets are extended along the large scale field, as is the case in quasi-2D magnetohydrodynamics \citep{Zank92}. The current sheet properties have a weaker dependence on the coordinate along the large scale field than on the coordinates in a plane perpendicular to that field.  I begin by assuming that one can define a ``domain'' which has a scale $\Lambda$ perpendicular to the large scale coronal magnetic field, and which contains a small integer number $N$ of current sheets. In the analysis which follows in Sections 3.1 and 3.2, I will assume that all current sheets are identical. The current sheet properties which are introduced are obviously to be understood as mean values from a distribution. Figure 1 illustrates such a domain.  The extent of the domain in the direction perpendicular to the plane defined by Figure 1 is $\Lambda_z = \mu \Lambda$, with $\mu > 1$. 
The current sheets have a width $L_c$, a thickness $t_c$, and an extension along the large scale field $Z_c > L_c$.   

The picture which has been drawn so far is consistent with the original view of \cite{Parker72}. It is also consistent with  results from studies of 2D MHD turbulence, which show that turbulent evolution results in the formation of isolated, intense sheets of current and vorticity.  The development which follows is based on results from \cite{Spangler99}, which contains an extensive bibliography to the literature where these ideas were developed earlier, most importantly \cite{Zank92}.  

In the case of current sheets which arise from 2D MHD turbulence, the number of positive and negative current sheets should be equal, and the expectation value of the current in an Amperian loop is zero.  The detection of net currents (via differential Faraday rotation) would then be interpreted as a statistical fluctuation of the total current about the zero expectation value. In what follows, I will refer to these as ``turbulent current sheets''.   

It is also possible that the physics of the corona selects current sheets with a preferred sign of the current density, at least for that portion of the corona which is probed in a Faraday rotation experiment.  This situation is referred to as that of ``deterministic current sheets'', and is discussed in Section 3.2.  Within this model, I  assume that the properties of the individual current sheets are essentially the same as in the turbulent model, but that there is a preference for one sign of current density.  It should be noted that the turbulent current sheet model is based on analytic and numerical solutions of the equations of 2D magnetohydrodynamics, whereas the deterministic model is plausible but  ad-hoc.  
\subsection{Heating from Turbulent Current Sheets}
I begin with the view that the current sheets arise as the evolution of 2D, or quasi-2D MHD turbulence \citep{Zank92,Spangler99}.  In this case, the domain size $\Lambda$ may be plausibly identified with the outer scale of the turbulence.  The Joule heating in each current sheet $\dot{\cal E}$ is 
\begin{equation}
\dot{\cal E}= \eta j^2 V = \eta j^2 L_c t_c Z_c
\end{equation}
where $V$ is the volume of a single sheet, given by $V=L_c t_c Z_c$. The Joule heating from all the current sheets in the domain $\dot{E}$ is then given by 
\begin{equation}
\dot{E}=N\dot{\cal E} = N\eta j^2 L_c t_c Z_c
\end{equation}
where $N$ is the number of current sheets per domain. The mean volumetric heating rate in the domain $\epsilon$, which is taken to be the overall volumetric heating rate, is 
\begin{equation}
\epsilon = \frac{\dot{E}}{V_D} = \frac{N \eta j^2 L_c t_c Z_c}{\Lambda^2 \Lambda_z}
\end{equation}
where $V_D=\Lambda^2 \Lambda_z$ is the volume of a domain.  
Using the fact that the current per sheet is $I=j L_c t_c$, we have 
\begin{equation}
\epsilon=\left( \frac{N \eta Z_c}{\Lambda^2 \Lambda_z}\right) \left[ \frac{I^2}{L_c t_c}\right]
\end{equation} 

The question now arises as to how to relate the current in an individual current sheet, $I$, with the total current $I_{obs}$ within the Amperian Loop. This relation will depend on the model for the current sheets.  For the remainder of this subsection, I will adopt the turbulence model in which there are, on  average, equal numbers of positive and negative current sheets, and statistical fluctuations are responsible for $I_{obs} \neq 0$.  

Let $N_T$ be the total number of current sheets within the Amperian Loop.  We then identify the measured current $I_{obs}$ with the rms fluctuation in the total current contained within the Loop, 
\begin{equation}
I_{obs} = \sqrt N_T I
\end{equation}
The total number $N_T$ is given by 
\begin{equation}
N_T=\frac{A}{\Lambda^2}N = \left( \frac{lS_{los}}{\Lambda^2}\right)N
\end{equation}   
where $A$ is the area of the Amperian Loop, $l$ is the spacing between the lines of sight, introduced in Section 1, and $S_{los}$ is the effective line-of-sight extent of the coronal plasma. Equation (8) is for the simplest case, in which the Amperian Loop is perpendicular to the large scale field.  In the general case, a cosine of an orientation angle would be introduced in the numerator.  This detail is ignored in the present discussion.  Substitution of equations (7) and (8) into equation (6) yields the volumetric heating rate in terms of the measured total current $I_{obs}$, 
\begin{equation}
\epsilon = \left[ \frac{\eta Z_c}{\Lambda_z L_c t_c} \right] \left( \frac{ I_{obs}^2}{lS_{los}} \right)
\end{equation}
As a final approximation, I assume that the extension of the domain and that of the current sheet along the large scale field direction  are described by the same anisotropy index $\mu$, $\Lambda_z=\mu\Lambda, Z_c = \mu L_c$.  Use of these relations gives us the basic expression for the average volumetric heating rate due to turbulent current sheets in terms of the observed parameter $I_{obs}$
\begin{equation}
\epsilon = \left( \frac{\eta}{\Lambda}\right) \left[ \frac{1}{t_c}\right] \left( \frac{I_{obs}^2}{lS_{los}}\right)
\end{equation}
This expression factors itself neatly into three terms, each contained within brackets.  The first is determined by the resistivity in the plasma and the domain properties.  The second is determined by properties of the current sheets, specifically their thickness.  The final term collects properties of the observations, such as the inferred total current and the parameters of the lines of sight.  
\subsection{Heating from Deterministic Current Sheets}
In this subsection, I consider the possibility that the current sheets are not entirely random, and that there may be some preference for one sign of the current density, probably determined by the polarity of the large-scale coronal field. The total net current from the Sun must obviously be zero.  However, it is possible for a net current to exist in a limited region probed by a radio remote sensing measurement. We assume that the properties of the individual current sheets can be described as previously, so that the equations of Section 3.1 up to, and including equation (6), are valid.  However, in the present case, there will be a different relationship between the total current $I_{obs}$ and the current of an individual sheet, $I$. If there is a preference for current sheets of one sign of the current density, we can write 
\begin{equation}
I_{obs} = \left( N_+ - N_- \right) I
\end{equation}
where $N_+$ is the number of sheets with positive current within the Amperian Loop, and $N_-$ is the number of sheets with negative current density. The individual sheet current $I$ is then taken as an absolute magnitude, with the sign of the current assumed in $N_+$ and $N_-$.  If we introduce probabilities that the current densities will be positive or negative by $N_+ = f_+ N_T, N_- = f_- N_T$, we have an expression for the current in a single current sheet, 
\begin{equation}
I = \frac{I_{obs}}{N_T(1-2f_-)}
\end{equation}
The total current sheet number $N_T$ is the same as that defined in equation (8).  Substitution of equation (12) into (6), and algebraic manipulation gives the volumetric heating rate in the case of ``deterministic'' current sheets
\begin{equation}
\epsilon = \left[ \frac{\eta \Lambda}{N}\right] \left( \frac{1}{t_c(1-2f_-)^2} \right) \left( \frac{I_{obs}^2}{l^2 S_{los}^2} \right)
\end{equation}
where the expression has again been factored into terms which contain, respectively, characteristics of the plasma, the current sheets, and the observations.  
\subsection{Comparison of the Expressions for the Heating Rate}
The expressions for the volumetric heating rate in the two models of the current sheets, equations (10) and (13) respectively, appear quite different in form, and it is natural to ask which is the larger for realistic input parameters.  In other words, given a measurement of $I_{obs}$, would greater Joule heating result if the current were distributed in a random set of turbulent current sheets as described in Section 3.1, or in a set of sheets with predominantly one sign of the current density, as discussed in Section 3.2.  

Let the heating rate expression for a turbulent set of current sheets as given in equation (10) be noted by $\epsilon_T$, and that due to a systematic set of sheets with preferentially one sign of the current density (equation (13))as $\epsilon_S$.  Equations (10) and (13) can be easily manipulated into the following form
\begin{equation}
\frac{\epsilon_S}{\epsilon_T} =  \left[ \frac{1}{N(1-2f_-)^2}\right] \left( \frac{\Lambda^2}{l S_{los}}\right) 
\end{equation}
If one assumes that the first term in square brackets on the right hand side of this equation is of order unity, then the relative heating rate depends on the ratio of the domain area to that of the Amperian loop. The precise value of this ratio depends on the circumstances of the observations, as well as the value of $\Lambda$.  An estimate of its value in the case of the observations of \cite{Spangler07} is given at the end of the next section.  In what follows, I discuss the case of turbulent current sheets, then briefly note that the conclusions would not be significantly different for the deterministic case.  
\subsection{Estimate of the Turbulent Heating Rate} 
Equation (10) is now used to estimate the coronal heating rate from turbulent current sheets.  The variables in the last term ($I_{obs},l,S_{los}$) are observational parameters and are known. The calculation will be carried out for the conditions characteristic of the large $\Delta RM$ event of August 16, 2003, in which the inferred current was $2.5 \times 10^9$ Amps. The line of sight to the radio source had a minimum heliocentric distance of $6.7 R_{\odot}$.  Coronal plasma properties characteristic of this distance will be used in the calculation below. A similar analysis at other times in the observations of \cite{Spangler07}, when there were only upper limits to the current, would obviously yield lower values for the Joule heating rate. The calculation also requires estimates of $\eta$, $\Lambda$, and $t_c$.  

\paragraph{Resistivity} For the resistivity $\eta$,  the Spitzer resistivity is used, which is based on Coulomb collisions of current-carrying electrons with ions and other electrons.  It is certain to be a drastic underestimate, in that the true resistivity is almost certainly determined by collisionless processes.  However, the Spitzer resistivity can be derived from fundamental principles, which is not true of other estimates, and it can serve as a lower limit to the true resistivity. An informal discussion of the possible role of collisionless processes in determining the resistivity is given in Section 4 below.  

The Spitzer resistivity is the reciprocal of the conductivity given by \cite{Gurnett05}
\begin{equation}
\sigma= \frac{32 \sqrt \pi \epsilon_0^2 (2 k_B T_e)^{3/2}}{\sqrt m_e e^2 ln \Lambda}
\end{equation} 
In this equation, and equation (16) below, $\Lambda$ stands for the Coulomb logarithm rather than the domain size as used otherwise.  The electron temperature is $T_e$.  All other terms in equation (15) have been defined, or are obvious fundamental physical constants.  

Equation (15) can be used to write the Spitzer resistivity in a ``suitable for observers'' form as \citep{Gurnett05}
\begin{equation}
\eta_S = 5.2 \times 10^{-5} \left( \frac{\ln \Lambda}{(k_B T_e)^{3/2}} \right) \mbox{ Ohm-m} 
\end{equation}
where the thermal energy $k_B T_e$ is now given in electron volts. For approximate coronal conditions I choose a value for the Coulomb logarithm of $\Lambda=25$ \citep{Krall73}. With an assumed coronal temperature of $2 \times 10^6$ K, appropriate for closed-field regions (electron thermal energy $k_B T = 172$ eV in equation (16)), the resistivity is $\eta_S=5.74 \times 10^{-7}$ Ohm-m, or about 35 times the resistivity of silver. 
\paragraph{Domain Size } I will take the domain size $\Lambda$ to be the outer scale of the turbulence in the relevant part of the corona.  There are two estimates in the literature for this outer scale.  The first is mean spacing between flux tubes which expand into the corona.  This estimate was introduced by \cite{Hollweg82}, and subsequently used by \cite{Mancuso99} and \cite{Cranmer05}.  
The formula used by \cite{Mancuso99} is 
\begin{equation}
\Lambda = \frac{1.37 \times 10^7}{\sqrt B(G)} \mbox{ meters}
\end{equation}
where $B(G)$ is the magnetic field strength in Gauss.  For the magnetic field in the corona, we use the recent estimate of \cite{Ingleby07} which was obtained from Faraday rotation measurements very similar to those of \cite{Spangler07}.  They found that the magnetic field could be represented by an inverse square dependence on the heliocentric distance, with a normalizing value of $\sim 0.050$ G at $r=5 R_{\odot}$.  At a heliocentric distance of 6.7 $R_{\odot}$, the estimated magnetic field is $2.78 \times 10^{-2}$ G, and the corresponding value of $\Lambda$ is $8.2 \times 10^7$ m.  In a more recent theoretical study of coronal heating and solar wind acceleration, \cite{Cranmer07} argue for a smaller value of the domain size (their parameter $L_{\perp}$ which serves as the outer scale of the turbulence), which physically corresponds to the diameter of the photospheric flux tubes rather than their separation.  Adopting the estimate for $\Lambda = L_{\perp}$ from \cite{Cranmer07} would reduce our value of $\Lambda$ by about a factor of 4 from the estimate of equation (17).  

A second estimate of the outer scale of coronal turbulence comes from power spectra of fluctuating Doppler shifts of a spacecraft transmitter \citep{Wohlmuth01,Efimov04}. These estimates, which result from measurements rather than plausible theoretical arguments, give outer scales from a few tenths of a solar radius to a solar radius or more at heliocentric distances of $5-10 R_{\odot}$.  The values reported by \cite{Wohlmuth01} and \cite{Efimov04} are several times larger than that given by equation (17). It is obvious that the factor of 4 smaller value for $\Lambda$ advocated by \cite{Cranmer07} is in more serious disagreement with the observational value of \cite{Wohlmuth01} and \cite{Efimov04}. A resolution of this matter would warrant a paper in its own right, but for the present work we use equation (17).  As may be seen from the heating rate expression in equation (10), lower values of the domain size $\Lambda$ generate higher values of the heating rate $\epsilon$.  
\paragraph{Current Sheet Thickness }  For the current sheet thickness $t_c$, I choose the ion inertial length $t_c = \frac{V_A}{\Omega_i}$ where $V_A$ is the Alfv\'{e}n speed, and $\Omega_i$ is the proton ion cyclotron frequency.  This would seem to be both plausible and a good lower limit to what the current sheet thickness can be.   Once again, equation (10) shows that use of a minimum plausible value for $t_c$ leads to an upper limit to the heating rate $\epsilon$.  To calculate the ion inertial length, the plasma density profile given by equation (6) of \cite{Spangler07} (based on radio propagation measurements of the corona) and the magnetic field model of \cite{Ingleby07} are used.  These yield the following formula for the estimated current sheet thickness
\begin{equation}
t_c = 1.0 \times 10^3 \left( \frac{R_0}{5}\right)^{1.2}  \mbox{ meters}
\end{equation}
where $R_0$ is the heliocentric distance in units of a solar radius. For $R_0 = 6.7 R_{\odot}$, $t_c = 1.4 \times 10^3$ m.  
\paragraph{Observational Parameters} The observed parameters in equation (10) are contained in the term in the third set of brackets.  The observed current $I_{obs} = 2.5 \times 10^9$ Amps and the separation of the lines of sight $l$ is 33,000 km \citep{Spangler07}.  There remains the value for the effective thickness of the plasma along the line of sight. I use the expression from \cite{Spangler02}
\begin{equation}
S_{los} = \frac{\pi}{2} R_0 R_{\odot}
\end{equation}

Use of the above parameters with an impact parameter $R_0 = 6.7$ gives a heating rate $\epsilon = 1.27 \times 10^{-16}$ Watts/m$^3$. To determine the significance of this number, I compare it to theoretical estimates of  \cite{Cranmer05} and \cite{Cranmer07}. These papers utilize cgs units, and report heating rates in power per unit mass. The volumetric heating rate given above is then $1.27 \times 10^{-15}$ ergs/cm$^3$/sec, and is converted to a heating rate/unit mass $q$
\begin{equation}
q = \epsilon/\rho = 3.0 \times 10^4 \mbox{ ergs/sec/gm}
\end{equation} 
where I have again used the power law density model of equation (6) of \cite{Spangler07} in obtaining the mass density at $r=6.7 R_{\odot}$. 

\cite{Cranmer07} calculate heating rates as a function of heliocentric distance.  Since their calculations are self-consistent, their heating rates may be considered to be those which are required by the observed heating of the corona and acceleration of the solar wind. Examination of Figure 8 of \cite{Cranmer07} shows that the heating rate per unit mass at a heliocentric distance of $\simeq 7 R_{\odot}$ is in the range of $10^{10} - 10^{11}$ ergs-sec$^{-1}$-gm$^{-1}$, depending on the assumed amplitude of the photospheric velocity fluctuations.  We therefore conclude that the heating rate given by equation (10) is lower than values which are required to account for coronal heating by at least a factor of $3 \times 10^5$, if the input parameters used here are valid.  If the outer scale to the turbulence is a factor of $\sim 4$ less than the value given by equation (17), as recommended by \cite{Cranmer07}, the ratio of mass heating rates would be about $10^5$.  In either case, this huge mismatch means that exercises with fine tuning the parameters in the model would be a fool's errand. It should be noted that the ratio $\frac{\Lambda^2}{l S_{los}}$ which appears in equation (14) is of the order of unity, within a factor of several either larger or smaller depending on the assumed outer scale of the turbulence. The conclusion on the magnitude of Joule heating would not be changed by adopting the non-turbulent current sheet model of Section 3.2.  

There are two possible conclusions to be drawn from the calculations of this section.  
\begin{enumerate}
\item In view of the large disparity between the calculated Joule heating rate and that which is required for a significant contribution to the thermodynamics of the corona, the currents which may have been observed are irrelevant for coronal heating. This argument would seem to be strengthened by the fact that I used the largest detected value of $I_{obs}$ from the two days of observation.  Other intervals would have provided smaller values for $I_{obs}$ or upper limits thereto, yielding smaller values of $\epsilon$.  
\item A more likely explanation, in my opinion, is that these current systems do play an important role in coronal heating, but that role is underestimated in the calculations presented here, because they are based on the Spitzer resistivity.  According to this viewpoint, the analysis of Sections 3.1 and 3.2 is valid, but a correct calculation would require an appropriate, and much larger value of the resistivity.  This is clearly speculation until it can be demonstrated that a much larger resistivity (by orders of magnitude) characterizes the coronal plasma at $5 R_{\odot} \leq r \leq 10 R_{\odot}$.   
\end{enumerate}

\section{Possible Enhancement of Resistivity in Current Sheets}
In this section, I consider the second of the possibilities listed immediately above, i.e. that the resistivity could be sufficiently enhanced in these coronal current sheets to make Joule heating a thermodynamically important process. An obvious way for this to happen is a plasma instability that produces high levels of fluctuating electric or magnetic fields, which scatter the current-carrying electrons and enhance the resistivity.  To assess this possibility, we need to examine the magnitude of the electron drift speed within the current sheets.

Equations (7) and (8) give the relationship between the observed current in the Amperian loop, $I_{obs}$, and the current in a single sheet, $I$.   Using these equations and the identity immediately before equation (6), we have for the current density in a single sheet
\begin{equation}
j = \frac{\Lambda I_{obs}}{t_c L_c \sqrt (l S_{los}) \sqrt N}
\end{equation} 
and for the electron drift speed
\begin{equation}
v_D = \frac{j}{e n} = \frac{\Lambda I_{obs}}{e n t_c L_c \sqrt (l S_{los}) \sqrt N} 
\end{equation}
where $e$ is the fundamental electric charge. To simplify equation (22) I adopt a set of plausible assumptions. I assume that the width of a current sheet will be some fraction of the domain size, $L_c = \beta \Lambda$ with $\beta$ probably having a value between 0.1 and 0.5.  
From Section 3, we already have estimates of other parameters (e.g. $t_c$, $L_c$, $n$) in equation (21) at the fiducial heliocentric distance of $6.7 R_{\odot}$.
This yields the following estimate for the electron drift speed, 
\begin{equation}
v_D = \frac{2.17 \times 10^6}{\beta \sqrt N} \mbox{ m/sec}
\end{equation}

For this expression to be meaningful in the context of plasma instabilities, we need to compare it with a characteristic plasma speed. An obvious choice is the electron thermal speed $v_{\theta} = \sqrt \frac{k_B T}{m_e}$ \citep{Nicholson83}. I use a value of $T_e = 2 \times 10^6$ K, which is characteristic of closed-magnetic-field regions in the corona.  Open field regions would have a lower temperature and lower thermal speed. We then have for the drift speed to thermal speed ratio
\begin{equation}
\frac{v_D}{v_{\theta}} = \frac{0.39}{\beta \sqrt N}
\end{equation}

As mentioned in the definition of $\beta$ immediately above, and the discussion of $N$ in Section 3, $\beta$ is a number which is probably less than unity, but not by a large factor, and $N$ is an integer which is probably larger than unity, but not much greater.  Their product should therefore be of order unity.  This calculation then suggests the plausibility of electron drift speeds of order the thermal speed in these current sheets.   If the drift speed is of the order of the electron thermal speed, it is also of order or larger than the ion acoustic speed.  

It is necessary to stress that this calculation has contained products of several parameters (such as $\Lambda, L_c$, etc) which are imperfectly known, so the net result presented here is similarly uncertain.  However, the conclusion of this section is that a current-driven instability, which would lead to high levels of fluctuating electric and magnetic fields, is a possibility.  

This calculation has also been carried out for the case of ``deterministic'' current sheets discussed in Section 3.2.  The details of the calculation are not presented here, but the final result is that the drift-to-thermal speed ratio is somewhat smaller (a factor of about 0.17), but not enough to alter the qualitative conclusion stated above. 

The possible existence of substantial electron drift speeds, comparable to the electron thermal speed, raises the possibility of an interesting observational diagnostic of such current sheets.  \cite{Spangler98} pointed out that an electron distribution carrying a current will have its distribution function distorted in the direction of current flow, and accordingly have a more populated tail than a distribution which carries no current.  This additional tail component makes the plasma more effective at collision excitation of ions to excited states whose energy above the ground state is a few times the electron thermal energy. \cite{Spangler98} suggested that turbulent current sheets might reveal themselves via enhanced emission line glow as the excited ions radiatively de-excite.  An important parameter determining the intensity of the line radiation is  
\begin{equation}
A_D \equiv \frac{m_e v_D^2}{2 k_B T}
\end{equation} 
This parameter is approximately equal to the square of the drift speed to thermal speed ratio.  When $A_D$ becomes of the order of a few tenths, the line emission can be substantially enhanced relative to the current-free value \citep[see Figure 9 of ][]{Spangler98}.  \cite{Spangler98} found that $A_D \ll 1$ for turbulence in the interstellar medium, so turbulent enhancement of emission line radiation probably does not occur there.  However, the results presented in this section suggest that this mechanism is much more likely to occur in the solar corona.  
\subsection{The consequences of current-driven instabilities}
A complete discussion of the consequences of a current-driven plasma instability for the resistivity  within coronal current sheets is beyond the scope of the present paper.  I will only briefly refer to some results in the literature which indicate the effect may be significant. 

The issue of instabilities due to high electron drift speeds was discussed in \cite{Spangler98}, who cited results from \cite{Drummond62}.  The remarks made there are still relevant to the present discussion. The summary of the work of \cite{Drummond62} presented in \cite{Spangler98} is that an electron drift speed greater than $\sim 0.12 v_{\theta}$ could be sufficient for excitation of obliquely-propagating electrostatic ion cyclotron waves.  

The role of a current-driven instability in enhanced resistivity was discussed by \cite{Chittenden95}. \cite{Chittenden95} developed a fluid theory to explain observations of laboratory Z-pinches which showed these structures to be larger than expected on the basis of theory with a Spitzer conductivity. The unexpectedly large size of Z pinches suggested that enhanced transport coefficients were present. \cite{Chittenden95} found that the electron drift speed exceeded the ion acoustic speed in the outer edges of the Z pinch.  His results (see Figure 3 of \cite{Chittenden95}) showed that the effective resistivity due to lower hybrid waves could exceed the Spitzer resistivity by 3 - 4 orders of magnitude.  This enhancement is approaching that needed for thermodynamic relevance of coronal currents, as discussed in Section 3.4. 
 
Enhanced resistivity has also been hypothesized to play an important role in magnetic reconnection, allowing reconnection to proceed at a faster rate and produce heating of the plasma.  \cite{Kulsrud05} discussed experiments showing the presence of magnetic fluctuations within the reconnection current sheet on the MRX experiment.  \cite{Kulsrud05} identify these fluctuations as obliquely-propagating waves arising due to a cross-field current which has a drift speed exceeding the Alfv\'{e}n speed.  \cite{Kulsrud05} found an enhancement of the resistivity, estimated from the wave force on the electrons, which exceeds the Spitzer resistivity by a factor of several.  The relatively modest enhancement of the resistivity relative to that required for coronal relevance, or the results discussed by \cite{Chittenden95}, can be attributed to the nearly collisional dynamics of the MRX experiment.  

The discussion in \cite{Kulsrud05} has recently been superceded by the results of \cite{Wang07,Wang08}, who now favor perpendicularly-propagating, unstable waves which nonlinearly couple to magnetosonic waves.  It is the magnetosonic waves which determine the resistivity.  This anomalous resistivity is higher than estimated in \cite{Kulsrud05}.  

The above-cited studies are not intended to correspond in detail to the case of Joule heating of the solar corona and the highly enhanced resistivity that would be required there.  The investigations of \cite{Chittenden95}, \cite{Kulsrud05} and \cite{Wang07,Wang08} are of importance in showing that current filaments with drift speeds comparable to or exceeding the ion acoustic speed can produce wave and turbulence fields that substantially enhance the resistivity. Current sheets or filaments with drift speeds approaching the electron thermal speed would be even more subject to instability, and to a wider range of unstable modes. 

A final point to be considered in this section is whether the resistivity in the coronal current sheets could plausibly reach levels necessary for important Joule heating.  The arguments in the previous paragraphs have shown that current-driven instabilities could quite plausibly be present, but could they enhance the resistivity by several orders of magnitude?  A very general expression for the resistivity is 
\begin{equation}
\eta = \frac{m_e \nu}{n e^2}
\end{equation} 
where $\nu$ is a collision frequency of some sort.  If collisionless scattering of electrons determines the resistivity, $\nu$ could be as large as the frequency of a high frequency plasma mode.  In what follows, I choose the lower hybrid frequency 
\begin{equation}
\nu_{LH} \simeq \frac{1}{2 \pi} \sqrt ( \Omega_i \Omega_e )
\end{equation}
In this identity, $\Omega_i$ and $\Omega_e$ are, respectively, the ion and electron cyclotron frequencies.  

I use the lower hybrid frequency as a proxy for a plasma mode frequency which is above the ion cyclotron frequency, and do not claim that the unstable waves in coronal current sheets are necessarily lower hybrid waves.  In support of this approach, \cite{Chittenden95} presents estimates of the anomalous collision frequency which are of order $\omega_{LH}$, with a multiplicative constant dependent on the drift speed to ion acoustic speed ratio \citep[see equation (2) of][]{Chittenden95}.

Using the coronal magnetic field model of \cite{Ingleby07}, we estimate a lower hybrid frequency  $\nu_{LH} = 1.82$ kHz at a heliocentric distance $r = 6.7 R_{\odot}$.  Substitution of this collision frequency into equation (25) (again using the same estimate of the electron density $n$ used in equation (22)) gives an anomalous resistivity of 6.27 Ohm-m.  This exceeds the Spitzer resistivity calculated following equation (16) by approximately 7 orders of magnitude.  This enhancement in the resistivity is comparable to, and in fact exceeds, the factor by which our calculated heating rate must be increased in order to be relevant for the thermodynamics of the corona \citep{Cranmer05,Cranmer07}.  This brief calculation then suggests that self-enhanced resistivity in coronal current sheets could lead to thermodynamically-relevant levels of Joule heating in the corona.  

\conclusions
\begin{enumerate}
\item Radioastronomical observations reported by \cite{Spangler07} are consistent with coronal currents flowing through Amperian loops defined by adjacent lines of sight to different components of a radio source. There are estimates of  currents of $2.5 \times 10^9$ and $2.3 \times 10^8$ Amperes, respectively, on two days.  Another interval of high quality data on one of the days yielded an upper limit to the differential Faraday rotation, and a corresponding upper limit to the current of $8 \times 10^8$ Amperes. These data are used as input for a calculation of Joule heating of the solar corona.  
\item Two models are developed to calculate the Joule heating associated with the observed currents. In both models, the current is envisioned as being in thin, intense current sheets stretched out along the large-scale coronal magnetic field.  The first views the sheets as arising in the evolution of quasi-2D magnetohydrodynamic turbulence.  The other assumes that current sheets will arise in the coronal plasma, and could show a preference for one sign of the current density.  These derivations are given in Sections 3.1 and 3.2, and provide the formulas for the volumetric heating rates given in equations (10) and (13).  
\item Use of these formulas, with observational data from \cite{Spangler07} and plausible independent coronal data, yield an estimated heating rate of $1.3 \times 10^{-16}$ Watts/m$^3$ ($1.3 \times 10^{-15}$ ergs/cm$^3$/sec in cgs units).  The corresponding heating rate per unit mass is $3.0 \times 10^4$ ergs/gm/sec.  This appears to be smaller than the level necessary to be significant for coronal heating by at least a factor of $3 \times 10^5$.  
\item The conclusion to be drawn from point (3) is that either these currents are irrelevant for coronal heating, or that the true resistivity in the corona exceeds the Spitzer value by several orders of magnitude.  Resolution of this matter obviously lies in a better understanding of the resistivity in a collisionless plasma. 
\item The same model used to estimate the volumetric and mass heating rates is also used to estimate the electron drift speed in the current sheets.  This drift speed could be comparable to the electron drift speed, and in excess of the ion acoustic speed.  Accordingly, current-driven instabilities might be present in these sheets, and the waves driven unstable by these currents might enhance the resistivity to significant levels. This contention is supported by works in the literature which have shown enhancement of resistivity by current-driven instabilities.  
\end{enumerate}

\begin{acknowledgements}
This work was supported at the University of Iowa by grant ATM03-54782 from the National Science Foundation. I thank Dr. Steven Cranmer of the Harvard-Smithsonian Center for Astrophysics for extensive commentary and suggestions on an earlier draft of this paper. I also thank Dr. S.P. Gary of the Los Alamos National Laboratory and Dr. Jack Scudder of the University of Iowa for discussions.  
\end{acknowledgements}


\begin{thebibliography}{}

\bibitem[Chittenden (1995)]{Chittenden95} Chittenden, J.P: The effect of lower hybrid instabilities on plasma confinement in fiber Z pinches, Phys. Plasmas~2, 1242, 1995
\bibitem[Cranmer and van Ballegooijen (2005)]{Cranmer05} Cranmer, S.R. and Ballegooijen, A.A.: On the generation, propagation, and reflection of Alfven waves from the solar photosphere to the distant heliosphere, Astrophys. J. Suppl. S.~156, 265, 2005
\bibitem[Cranmer et al (2007)]{Cranmer07} Cranmer, S.R., van Ballegooijen, A.A., and Edgar, R.J.: Self-consistent coronal heating and solar wind acceleration from anisotropic magnetohydrodynamic turbulence, Astrophys. J. Suppl. S.~171, 520, 2007
\bibitem[Drummond and Rosenbluth (1962)]{Drummond62} Drummond, W.E. and Rosenbluth, M.N.: Anomalous diffusion arising from microinstabilities in a plasma, Phys. Fluids~5, 1507, 1962
\bibitem[Efimov et al (2004)]{Efimov04} Efimov, A.I., Bird, M.K., Chashei, I.V., and Samoznaev, L.N.: Outer scale of solar wind turbulence deduced from two-way coronal radio sounding experiments, Adv. Space Res. 33, 701, 2004
\bibitem[Gudiksen and Nordlund (2005)]{Gudiksen05} Gudiksen, B.V. and Nordlund, \AA: An Ab Initio approach to the solar coronal heating problem, Astrophys. J.~618, 1020, 2005; {\em Erratum: } Astrophys. J.~623, 600
\bibitem[Gurnett and Bhattacharjee (2005)]{Gurnett05} Gurnett, D.A. and Bhattacharjee, A.: Introduction to Plasma Physics, Cambridge University Press, 2005, p430
\bibitem[Hollweg et al (1982)]{Hollweg82} Hollweg, J.V., Bird, M.K., Volland, H., Edenhofer, P., Stelzried, C.T., and Seidel, B.L.: Possible evidence for coronal Alfven waves, J. Geophys. Res.~87, 1, 1982
\bibitem[Ingleby et al (2007)]{Ingleby07} Ingleby, L.D., Spangler, S.R., and Whiting, C.A.: Probing the large-scale plasma structure of the solar corona with Faraday rotation measurements, Astrophys. J.~668, 520, 2007
\bibitem[Krall and Trivelpiece (1973)]{Krall73} Krall, N.A. and Trivelpiece, A.W.: Principles of Plasma Physics, McGraw-Hill, 1973, p294
\bibitem[Kulsrud et al (2005)]{Kulsrud05} Kulsrud, R., Ji, H., Fox, W., and Yamada, M.: An electromagnetic drift instability in the magnetic reconnection experiment and its importance for magnetic reconnection, Phys. Plasmas~12, 082301, 2005
\bibitem[Mancuso and Spangler (1999)]{Mancuso99} Mancuso, S. and Spangler, S.R.: Coronal Faraday rotation observations: measurements and limits on plasma inhomogeneities, Astrophys. J.~525, 195, 1999
\bibitem[Nicholson (1983)]{Nicholson83} Nicholson, D.R.: Introduction to Plasma Theory, John Wiley and Sons, 1983, p4
\bibitem[Parker (1972)]{Parker72} Parker, E.N.: Topological dissipation and the small-scale fields in turbulent gases, Astrophys. J.~174, 499, 1972
\bibitem[Peter et al (2006)]{Peter06} Peter, H., Gudiksen, B.V., and Nordlund, \AA.: Forward modeling of the corona of the Sun and solar-like stars: from a three-dimensional magnetohydrodynamic model to synthetic extreme-ultraviolet spectra, Astrophys. J.~638, 1086, 2006
\bibitem[Sakurai and Spangler (1994)]{Sakurai94} Sakurai, T. and Spangler, S.R.: The study of coronal plasma structures and fluctuations with Faraday rotation measurements, Astrophys. J.~434, 773, 1994 
\bibitem[Spangler (1998)]{Spangler98} Spangler, S.R.: Magnetohydrodynamic turbulence and enhanced atomic processes in astrophysical plasmas, Phys. Plasmas~5, 3006, 1998
\bibitem[Spangler (1999)]{Spangler99} Spangler, S.R.: Two dimensional magnetohydrodynamics and interstellar plasma turbulence, Astrophys. J.~522,879, 1999
\bibitem[Spangler (2002)]{Spangler02} Spangler, S.R.: The amplitude of magnetohydrodynamic turbulence in the inner solar wind, Astrophys. J.~576,997, 2002
\bibitem[Spangler (2007)]{Spangler07} Spangler, S.R.: A technique for measuring electrical currents in the solar corona, Astrophys. J.~670, 841, 2007
\bibitem[Wang, Kulsrud, and Ji (2007)]{Wang07} Wang, Y, Kulsrud, R., and Ji, H.: New results on the perpendicularly-propagating electromagnetic lower hybrid drift instability (LHDI), 49th Meeting of the APS Division of Plasma Physics, paper TP8.012, 2007
\bibitem[Wang, Kulsrud, and Ji (2008)]{Wang08} Wang, Y, Kulsrud, R., and Ji, H.: Nonlinear mode coupling calculation on the perpendicularly-propagating electromagnetic instabilities in the MRX, 50th Meeting of the APS Division of Plasma Physics, paper GP6.022, 2008
\bibitem[Wohlmuth et al (2001)]{Wohlmuth01} Wohlmuth, R., Plettemeier, D., Edenhofer, P., Bird, M.K., Efimov, A.I., Samoznaev, L.N., and Chashei, I.: Radio frequency fluctuation spectra during the solar conjunctions of the Ulysses and Galileo spacecraft, Space Sci. Rev.~97, 9, 2001
\bibitem[Zank and Matthaeus (1992)]{Zank92} Zank, G.P. and Matthaeus, W.H.: The equations of reduced magnetohydrodynamics, J. Plasm. Phys.~48, 85, 1992
\end{thebibliography}
\end{document}